\documentstyle[12pt,aasms4,epsf]{article}

\def\tempest%
{\begin{array}{ccc}
1 & 1 & 1 \\
1 & 1 & 1 \\
4 & 3 & 8
\end{array}}

\begin{document}

\title{The Effect of Bright Lenses on the Astrometric \\
       Observations of Gravitational Microlensing Events}
\bigskip
\bigskip

\author{Youngjin Jeong}
\author{Cheongho Han}
\author{Sung-Hong Park}
\bigskip
\affil{Department of Astronomy \& Space Science, \\
       Chungbuk National University, Cheongju, Korea 361-763 \\
       jeongyj@astro-3.chungbuk.ac.kr, \\
       cheongho@astro-3.chungbuk.ac.kr, \\
       parksh@astro-3.chungbuk.ac.kr}
\bigskip
\authoremail{cheongho@astro-3.chungbuk.ac.kr}
\authoremail{jeongyj@astro-3.chungbuk.ac.kr}
\authoremail{parksh@astro-3.chungbuk.ac.kr}
\bigskip

\begin{abstract}
   In current microlensing experiments, the information about 
the physical parameters of individual lenses are obtained from 
the Einstein timescales.  However,  the nature of MACHOs is still 
very uncertain despite the large number of detected events.
This uncertainty is mainly due to the degeneracy of the lens parameters 
in the measured Einstein timescales.  The degeneracy can be 
lifted in a general fashion if the angular Einstein ring radius 
$\theta_{\rm E}$, and thus the MACHO proper motion, can be measured 
by conducting accurate astrometric measurements of  
centroid displacement in the source star image.

   In this paper, we analyze the influence of bright lenses on
the astrometric measurements of the centroid displacement and
investigate this effect on the determination of $\theta_{\rm E}$.
We find that if an event is caused by a bright lens, the centroid
displacement is distorted by the flux of the lens and resulting
astrometric ellipse becomes rounder and smaller with increasing lens
brightness, causing an incorrect determination of the angular Einstein 
ring radius.  
A lens-blended event cannot be distinguished from a dark lens event
from the trajectory of the measured centroid displacements alone
because both events have elliptical trajectories: the degeneracy
between dark and bright lens events.  We also find that
even with information from the analysis of the light curve of the 
event it will still be difficult to resolve the degeneracy caused
by the bright lens because the light curve is also affected by lens 
blending.

\end{abstract}

\vskip25mm
\keywords{gravitational lensing -- dark matter -- astrometry}

\centerline{submitted to {\it The Astrophysical Journal}: July 16, 1998}
\centerline{Preprint: CNU-A\&SS-06/98}
\clearpage

\section{Introduction}

   Searches for gravitational microlensing events by monitoring stars
located in the Large Magellanic Cloud (LMC) and the Galactic bulge
are being carried out and $\sim 300$ events have been detected
(Alcock et al.\ 1997a, 1997b; Ansari et al.\ 1996; Udalski et al.\ 1997;
Alard \& Guibert 1997).\markcite{alcock1997, ansari1996, udalski1997,
alard1997}  All of the information from current microlensing experiments 
is obtained from the Einstein timescale, $t_{\rm E}$.  
One can obtain information about individual lenses because the Einstein 
timescale is related to the physical parameters of the lens by
$$
t_{\rm E} = {r_{\rm E} \over v}; \qquad
r_{\rm E} = \left( {4GM\over c^2}{D_{ol}D_{ls}\over D_{os}}\right)^{1/2},
\eqno(1)
$$
where $r_{\rm E}$ is the physical size of the Einstein ring, $M$ is the 
mass of the lens, $v$ is the lens-source transverse speed, $D_{ol}$, 
$D_{ls}$, and $D_{os}$ are the separations between the observer, lens, 
and source star.  In addition, one can obtain information about 
galactic structure and the halo MACHO fraction from the measured 
optical depth, which is directly proportional to the sum of the 
timescales of individual events. 

   However, the Einstein timescale alone cannot fully constrain the physical 
parameters of individual lenses.  This is because the Einstein timescale 
depends on a combination of the physical parameters of the lens, i.e., $M$, 
$v$, and $D_{ol}$.  As a result, the nature of MACHOs is still very 
uncertain despite the large number of detected events.
Therefore it is very important to devise a general method that 
can resolve the degeneracy in the lens parameters. 

   In general, the problem of breaking the lens parameter degeneracy has
been approached in two different ways.  The first approach is through 
measuring the lens parallax (Gould 1994b).\markcite{gould1994b}  
If the parallax of an event is measured, one can obtain the projected speed 
$\tilde{v}=(D_{ol}/D_{ls})v$.  The second approach is through measuring 
the lens proper motion, $\mu=\theta_{\rm E}/t_{\rm E}$, where 
$\theta_{\rm E}=r_{\rm E}/D_{ol}$ is the angular Einstein ring radius.
When both the values of $\tilde{v}$ and $\mu$ are determined, one can 
completely break the degeneracy and obtain the physical parameters
of individual lenses by
$$
\cases{
M = (c^2/4G)t_{\rm E}^2\tilde{v}\mu,\cr
D_{ol} = D_{os}(\mu D_{os}/\tilde{v} +1)^{-1},\cr
v = [\tilde{v}^{-1}+(\mu D_{os})^{-1}]^{-1}.
}
\eqno(2)
$$

   Many methods to measure the lens parallax and the proper motion 
have been proposed.  The parallaxes of lensing events can be measured in 
a general fashion from the simultaneous observations of the event from the 
ground and space (Gould 1994b, 1995; Han \& Gould 1995; 
Boutreux \& Gould 1997; Gaudi \& Gould 1997).\markcite{gould1994b, 
gould1995, han1995, boutreux1997, gaudi1997}
One way to measure the lens proper motion is by utilizing the finite
source size effect of lensing events, which can be measured
photometrically (Gould 1994a; Nemiroff \& Wickramasinghe 1994; 
Witt \& Mao 1994; Witt 1995; Peng 1997)\markcite{gould1994a, nemiroff1994,
witt1994, witt1995, peng1997}, spectroscopically (Maoz \& Gould 1994; 
Loeb \& Sasselov 1995)\markcite{maoz1994, loeb1995}, and astrometrically 
(Mao \& Witt 1998)\markcite{mao1998}.  Proper motions can also be measured 
by using lunar occultation (Han, Narayanan, \& Gould 1996)\markcite{han1996} 
and by analyzing the light curves of binary source events 
(Han \& Gould 1997).\markcite{han1997} However, unlike the measurement of 
lens parallax, measurement of $\mu$ using the above methods is 
only possible for a few rare cases, such as very close lens-source 
encounter events.

   A much more general way of measuring the proper motion is by conducting an 
accurate astrometric measurement of the separation between the two source 
images during the event (Miyamoto \& Yoshii 1995).\markcite{miyamoto199}
This suggestion, however, is implausible because the expected 
separation for typical values of lens mass and distance is too small
for direct observation of individual images.  However, several 
planned interferometric observation missions, e.g., the 
{\it Space Interferometry Mission} 
(Allen, Shao, \& Peterson 1997, hereafter SIM)\markcite{allen1997}, 
can be used to measure the astrometric displacements in the light 
centroid caused by gravitational microlensing in events detected 
photometrically from the ground (Walker 1995; 
H\o\hskip-1pt g, Novikov \& Polnarev 1995; Paczy\'nski 1998; 
Boden, Shao, \& Van Buren 1998).\markcite{walker1995, hog1995
paczynski1998, boden1998}  During the event, the apparent position of 
the lensed source star traces out an ellipse (`astrometric ellipse'), 
whose ellipticity 
depends on the lens-source impact parameter.  The proper motion can be 
measured because the size of the astrometric ellipse is directly 
proportional to the Einstein ring radius (see \S\ 2).  However, except 
for a brief discussion of bright lenses by Boden et al.\ 
(1997),\markcite{boden1997} all previous studies are based on the 
assumption of a dark lens.

   In this paper, we analyze the influence of bright lenses on 
the astrometric measurements of the centroid displacement and
investigate this effect on the determination of $\theta_{\rm E}$.
We find that if an event is caused by a bright lens, the centroid 
displacement is distorted by the flux of the lens and the resulting 
astrometric ellipse becomes rounder and smaller with increasing lens
brightness, causing an incorrect determination of the angular 
Einstein ring radius.
A lens-blended event cannot be distinguished from a dark lens event
from the trajectory of the measured centroid displacements alone
because both events have elliptical trajectories: the degeneracy
between dark and bright lens events.  We also find that even with 
information from the analysis of the light curve of the event it will 
still be difficult to resolve the degeneracy caused by the bright lens 
because the light curve is also affected by lens blending.

\section{Astrometric Shift of the Source Image Centroid: Dark Lens Case} 

   When a source star is microlensed, it is 
split into two images.  For a source with a separation $u$ from a lens, 
the images are seen at 
$$
\theta_{I,\pm} = 
{1\over 2}\left[ u\pm (u^2 +4)^{1/2}\right] \theta_{\rm E},
\eqno(3)
$$
with individual amplifications of 
$$
A_\pm = {1\over 2}\left[ {u^2+2\over u(u^2+4)^{1/2}}\pm 1\right].
\eqno(4)
$$
The lens-source separation in units of the angular Einstein ring 
radius is related to the lensing parameters by
$$
u = \left[ \beta^2 + {\cal T}^2 \right]^{1/2};\qquad
{\cal T}= {t-t_0\over t_{\rm E}},
\eqno(5)
$$
where $\beta$ is the impact parameter and $t_0$ is the time of maximum 
amplification.  By symmetry, the images with respect to the lens lie 
at the same azimuth as the source, one on the same and the other 
on the opposite side of the source.  Under the assumption that the lens 
is dark, the centroid of the image pair is defined by
$$
\theta_c = {A_+\theta_{I,+}+A_-\theta_{I,-} \over A_+ + A_-}
         = {1\over 2} \left[ {u(u^2+4)\over u^2+2}+u\right]\theta_{\rm E}.
\eqno(6)
$$
Then the location of the image centroid relative to the source 
star is determined by
$$
\vec{\delta\theta} = {\theta_{\rm E} \over u^2+2}({\cal T}\hat{\bf x} + 
	             \beta\hat{\bf y}),
\eqno(7)
$$
where the $x$ and $y$ axes represent the directions which are
parallel and normal to the lens-source transverse motion.

   If we let $x=\delta\theta_{c,x}$ and $y=\delta\theta_{c,y}-b_0$,
$b_0 = \beta/2(\beta^2+2)\theta_{\rm E}$, the coordinates are related by
$$
x^2 + {y^2\over q_0^2} = a_0^2,
\eqno(8)
$$
where $a_0 = \theta_{\rm E} / 2(\beta^2+2)^{1/2}$
and $q_0 = b_0/a_0 = \beta / (\beta^2+2)^{1/2}$.
Therefore, the trajectory of the apparent source star image centroid 
traces out an ellipse during the event, the `astrometric ellipse'.

   Once the astrometric ellipse is measured, one can determine the impact 
parameter of the event because the axis ratio of the ellipse is related 
to the impact parameter.  What makes the measurement of astrometric 
shift useful lies in the fact that the size of the astrometric ellipse 
is directly proportional to the angular Einstein ring radius.  
Therefore, one can uniquely determine the proper motion by fitting the 
observed astrometric shifts to the theoretical trajectories in equations 
(7) combined with the timescale which is determined from the photometric 
observation of the event.  In Figure 1, we present the astrometric 
ellipses for various values of $\beta$ (upper panel) and the changes in
the semimajor axis (middle panel) and in the axis ratio (lower panel) of 
the ellipse as a function of the impact parameter.

\section{Astrometric Shift with Bright Lenses}

   If a lensing event is caused by a bright lens, on the other hand, 
the centroid shift is distorted by the flux of the lens: 
`astrometric lens blending.'  The centroid displacement, which is 
measured with respect to a new reference position for the unlensed 
lens-source centroid by subtracting the relative motion of the 
center with respect to nearby stars, for 
the lens-blended event becomes
$$
\vec{\delta\theta}_c = {\cal D}
	{\theta_{\rm E} \over u^2+2}({\cal T}\hat{\bf x} +
	\beta\hat{\bf y}).
\eqno(9)
$$
The distortion factor caused by lens blending is related to the flux 
ratio between the lens and the source star before gravitational lensing 
amplification by
$$
{\cal D}\left( u,{\ell_L\over\ell_S}\right) =
\left\{ 
1+(\ell_L/\ell_S)+(\ell_L/\ell_S)\left[ (u^2+2)-u(u^2+4)^{1/2}\right]
\over
\left[ 1+(\ell_L/\ell_S)\right]\left[ 1+(\ell_L/\ell_S)
u(u^2+4)^{1/2}/(u^2+2)\right]
\right\}.
\eqno(10)
$$

   The new trajectory is also an ellipse, but with different shape and
size from that of an event caused by a dark lens.  In the upper panel 
of Figure 2, we present the astrometric ellipses for various values of 
$\delta m$ for an example bright lens event with $\beta=0.5$, where 
$\delta m=m_L-m_S$ is the apparent magnitude difference between the 
lens and source.
By comparing the shapes and sizes of these ellipses for different
values of $\delta m$, one finds two trends: First, as the lens/source 
flux ratio increases, the astrometric ellipse becomes rounder.
The axis ratios are $q(\delta m)=0.388$, 0.446, 0.532, and 0.624 
for $\delta m=2$, 1, 0, and $-1$, respectively, compared to the value of 
$q_0=0.333$ for the dark lens case.  Second, the size of the ellipse 
(measured by the size of the semimajor axis) decreases with increasing 
lens brightness.  The semimajor axes of the ellipses in units of 
$\theta_{\rm E}$ when $\delta m=2$, 1, 0, and $-1$ are 
$a(\delta m)=0.312$, 0.284, 0.230, and 0.153, 
respectively, compared to the value of $a_0=0.333$ for the dark lens case.
In the lower panel of Figure 3, we present the changes in the axis ratio
($q/q_0$: deformation factor) and the semimajor axis ($a/a_0$: contraction
factor) of the astrometric ellipse as a function of $\delta m$.
Note that unlike the semimajor axis, the
semiminor axis, $b=aq$, does not monotonically decrease with increasing
lens brightness. It increases when $\delta m$ is large, but then decreases 
as the lens brightness increases further.  This transient increase in $b$ 
arises because the ellipse becomes both rounder and smaller.

\section{Determination of $\theta_{\rm E}$ for Lens-Blended Events} 

   A lens-blended event cannot be distinguished from a dark lens event
just from the trajectory of the measured centroid displacements because both
events have the same elliptical trajectories: the degeneracy between 
dark and bright lens events.   If the centroid shifts of a bright 
lens event is fit to a model for a dark lens event not knowing the event 
is affected by the lens brightness, 
the size of the angular Einstein ring radius will be underestimated.  
Since the size of the angular Einstein ring radius is directly proportional 
to the size of the astrometric ellipse, the ratio of the observed to the true 
size of the angular Einstein ring radius is determined by 
$$
{\theta_{\rm E,obs} \over \theta_{\rm E,true}} = 
{a_{\rm obs}\over a_{\rm true}} = 
{a(\beta_0)\over a_0(\beta_0)}
{a_0(\beta_0)\over a_0(\beta_{\rm obs})}
= {a(\beta_0)\over a_0(\beta_{\rm obs})},
\eqno(11)
$$
where the term $a(\beta_0)/ a_0(\beta_0)$ represents the contraction factor 
of the semimajor axis and the term $a_0(\beta_0)/ a_0(\beta_{\rm obs})$ is
included because $\theta_{\rm E}$ is determined not from $\beta_0$ but
from $\beta_{\rm obs}$, which is the impact parameter determined from the
shape of the observed astrometric ellipse.
   In Table 1, we list the values of best-fitting $\beta_{\rm obs}$ and 
$\theta_{\rm E,obs}/\theta_{\rm E,true}$ which are determined from the 
dark lens fit to the observed astrometric centroid displacements 
of an example event with $\beta_0=0.5$ when it is blended with
various lens brightnesses.
From the table, one finds that increasing the lens brightness causes  
$\beta_{\rm obs}$ to increase and the determined values of 
$\theta_{\rm E,obs}$ to decrease compare to their true values.

  An interesting fact to note is that although one obtains incorrect 
values of $\theta_{\rm E}$ if lens-blending is not considered, the 
determined proper motions of events are relatively less affected by 
lens blending.
This is because not only the astrometrically determined value of
$\theta_{\rm E}$, but also the photometrically determined $t_{\rm E}$
is underestimated due to lens blending.
When the light curve of an event is blended with a bright lens, the 
determined timescale is reduced by a factor
$$
{t_{\rm E,obs}\over t_{\rm E,true}} =
\left[ 2(1-A_{\rm min}^{-2})^{1/2} -2  \right]^{1/2};\qquad
A_{\rm min}=0.34(1+\ell_L/\ell_S)+1
\eqno(12)
$$
(Nemiroff 1997; Han 1998).\markcite{nemiroff1997, han1998}
In Table 1, we list the ratios of the photometrically determined 
timescale and resulting proper motion to their true values
for events blended by lenses with various brightnesses.
One finds that despite substantial changes in the values of both
$\theta_{\rm E}$ and $t_{\rm E}$, the determined proper motions
do not deviate seriously from the true values.

\section{Discussion}

   In the previous section, we showed that if an event is caused by 
a bright lens, the centroid displacement is distorted by the flux of 
the lens and the resulting astrometric ellipse becomes rounder and 
smaller with increasing lens brightnesses, causing an incorrect
 determination of $\theta_{\rm E}$.  
However, since the astrometric measurements of the centroid displacements 
will be conducted for events detected photometrically, one might think 
that it would be possible to correct the astrometric lens-blending effect
with additional information from the analysis of the light curves.
If the correct value of the impact parameter 
is obtained from the light curve, it will be possible to determine the 
lens/source flux ratio from the comparison of $\beta_{\rm obs}$ and 
$\beta_0$.  Once $\ell_L/\ell_S$ is determined, 
it will be explicit to obtain the correct value of $\theta_{\rm E}$.

   However, we find that it will still be difficult to correct 
the lens-blending effect even with photometric information because
light curve itself is also affected by lens blending: photometric
lens blending.  
To demonstrate this we simulate lens-blended events toward both the 
Galactic bulge and LMC under realistic observational conditions.  
We determine the uncertainty of the photometrically recovered $\beta$ 
by conducting both bright and dark lens fit to the simulated event 
light curve.  The result of the fit is obtained by computing $\chi^2$, i.e., 
$$
\chi^2 = 
{1\over N_{\rm obs}-N_{\rm par}}
\left[
\sum_{i=1}^{N_{\rm obs}} A_{\rm obs}(t_i)-A(t_i)
\over
p A(t_i)
\right]^2,
\eqno(13)
$$
where $N_{\rm obs}$ is the number of measurements, $p$ is 
the photometric precision, and $N_{\rm par}$ is the number of parameters, 
which is 5 ($t$, $t_0$, $t_{\rm E}$, $\beta$, and 
$\ell_L/\ell_S$).
The light curve with lens blending is represented by
$$
A = {A_{\rm abs} + \ell_L/\ell_S \over 1+ \ell_L/\ell_S},
\eqno(14)
$$
where $A_{\rm abs}=A_{+}+A_{-}=(u^2+2)/u(u^2+4)^{1/2}$ 
is the absolute amplification of the event.
The events are assumed to be observed with a frequency of 
$2\ {\rm times}\ {\rm day}^{-1}$ during $-3t_{\rm E}\leq t\leq 3t_{\rm E}$.
For the values of the Einstein timescales, we adopt 
$t_{\rm E}=20\ {\rm days}$ and 40 days for the Galactic bulge and 
LMC events, respectively, based on the most frequent values of 
actual events toward each direction
(Alcock et al.\ 1997a, 1997b).\markcite{alcock1997a, alcock1997b}
The photometric precision is assumed to be $p=2\%$ for the 
Galactic bulge events, while we assume $p=5\%$ for the LMC events
as the source stars are faint.  The uncertainties for both fits 
are determined by the $1\sigma$ level, which is equivalent to $\chi^2=1$.

   In Table 2, we list the uncertainties in the photometrically determined
impact parameters for bright lens events by both dark and bright lens fits. 
From the table, one finds the best-fitting impact parameter from the
astrometric dark lens fit, $\beta_{\rm obs}$ listed in Table 1, 
lies within the uncertainty range of $\beta$ obtained from the
photometric bright lens fit.  Although the best-fitting impact parameter 
from the astrometric dark lens fit is just beyond the uncertainty range of 
$\beta$ determined from the photometric dark lens fit, it will be 
very difficult to distinguish the two values considering the uncertainties 
accompanied to $\beta_{\rm obs}$.  This implies that lens-blending effect 
cannot be detected  from the simple comparison of the impact parameters 
which are determined both astrometrically and photometrically.
In addition, one finds that the best-fitting $\beta$ from the photometric
dark lens fit lies within the uncertainty range of $\beta$ obtained 
from the photometric lens blending fit, implying that lens blending 
cannot also be easily detected photometrically.

\acknowledgements
We would like to thank to P.\ Martini for careful reading the manuscript.

\clearpage

\begin{center}
\bigskip
\bigskip
\centerline{\small {TABLE 1}}
\smallskip
\centerline{\small {\sc Lens-Blending Effect on the Determination 
of Lens Parameters}} 
\smallskip
\begin{tabular}{ccccc}
\hline
\hline
\multicolumn{1}{c}{} &
\multicolumn{3}{c}{dark lens fit} &
\multicolumn{1}{c}{} \\
\multicolumn{1}{c}{$\delta m$} &
\multicolumn{2}{c}{astrometric} &
\multicolumn{1}{c}{photometric} &
\multicolumn{1}{c}{$\mu_{\rm obs}/\mu_{\rm true}$} \\
\multicolumn{1}{c}{(mag)} &
\multicolumn{1}{c}{$\beta_{\rm obs}$} &
\multicolumn{1}{c}{$\theta_{\rm E,obs}/\theta_{\rm E,true}$} &
\multicolumn{1}{c}{$t_{\rm E,obs}/t_{\rm E,true}$} &
\multicolumn{1}{c}{} \\
\hline
$\infty$ & 0.500  & 1.000 & 1.000 & 1.000 \\
  2      & 0.596  & 0.965 & 0.933 & 1.034 \\
  1      & 0.705  & 0.897 & 0.849 & 1.057 \\
  0      & 0.889  & 0.769 & 0.699 & 1.100 \\
 $-1$    & 1.129  & 0.561 & 0.497 & 1.129 \\
\hline
\end{tabular}
\end{center}
\smallskip
\noindent
{\footnotesize \qquad NOTE.---  
The values of best-fitting impact parameter $\beta_{\rm obs}$ and the 
ratio between the observed and true angular Einstein ring radius 
$\theta_{\rm E,obs}/\theta_{\rm E,true}$ which are determined from
the dark lens fit to the observed astrometric centroid displacements 
of an example event with $\beta_0=0.5$ when it is blended by 
the flux of lenses with various brightnesses.  One finds that 
increasing the lens brightness causes $\beta_{\rm obs}$ to increase 
and the determined values of $\theta_{\rm E,obs}$ to decrease compared 
to its true values.  Also listed are the ratios of the photometrically 
determined timescale and resulting proper motion to their true values.
One finds that despite substantial changes in the values of both 
$\theta_{\rm E}$ and $t_{\rm E}$, the determined proper motion
do not deviate seriously from the true values.
   }

\begin{center}
\bigskip
\bigskip
\centerline{\small {TABLE 2}}
\smallskip
\centerline{\small {\sc The Uncertainties of Photometrically 
Determined Values of $\beta$}} 
\smallskip
\begin{tabular}{ccccc}
\hline
\hline
\multicolumn{1}{c}{} &
\multicolumn{2}{c}{bright lens fit} &
\multicolumn{2}{c}{dark lens fit} \\
\multicolumn{1}{c}{$\delta m$} &
\multicolumn{2}{c}{$\beta+\Delta\beta$} &
\multicolumn{2}{c}{$\beta+\Delta\beta$} \\
\multicolumn{1}{c}{(mag)} &
\multicolumn{1}{c}{bulge events} &
\multicolumn{1}{c}{LMC events} &
\multicolumn{1}{c}{bulge events} &
\multicolumn{1}{c}{LMC events} \\
\hline
  2    & $0.5\pm 0.11$  & $0.5\pm 0.30$ & $0.55\pm 0.004$ & $0.55\pm 0.11$ \\
  1    & $0.5\pm 0.15$  & $0.5\pm 0.35$ & $0.63\pm 0.005$ & $0.63\pm 0.14$ \\
  0    & $0.5\pm 0.28$  & $0.5\pm 0.47$ & $0.77\pm 0.02$  & $0.77\pm 0.19$ \\
 $-1$  & $0.5\pm 0.55$  & $0.5\pm 0.70$ & $1.04\pm 0.06$  & $1.04\pm 0.36$ \\
\hline
\end{tabular}
\end{center}
\smallskip
\noindent
{\footnotesize \qquad NOTE.---
The uncertainties of the photometrically determined values of the 
impact parameter for events blended by lenses with various
brightnesses.  The uncertainties are determined by conducting both
bright and dark lens fits to the simulated event light curves 
constructed under reasonable observational conditions.
  }

\begin{figure}[t]
\centerline{\hbox{\epsfysize=16truecm \epsfbox{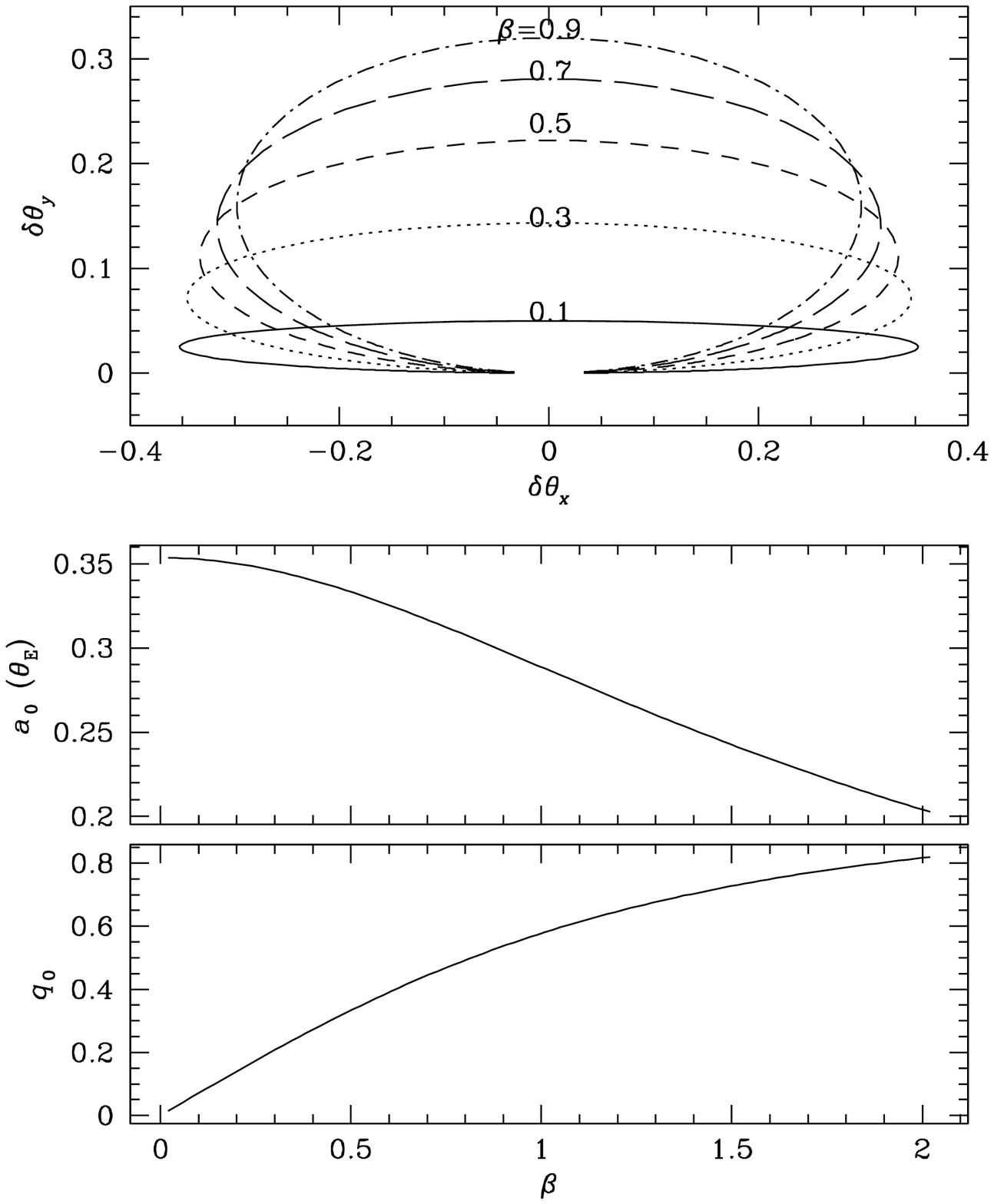}}}
\noindent
{\small {\bf Figure 1:}\
Upper panel:  
	The trajectories of the source image centroid displacements of 
	an event caused by a dark lens for various values of $\beta$.  
	Here, $x$ and $y$ axes represent the directions which are parallel 
	and normal to the lens-source transverse motion, respectively.
Middle and lower panels:
        The changes of semimajor axis (middle panel) and the axis ratio 
	(lower panel) of the astrometric ellipse as a function 
	of impact parameter.  
}
\end{figure}

\begin{figure}[t]
\centerline{\hbox{\epsfysize=16truecm \epsfbox{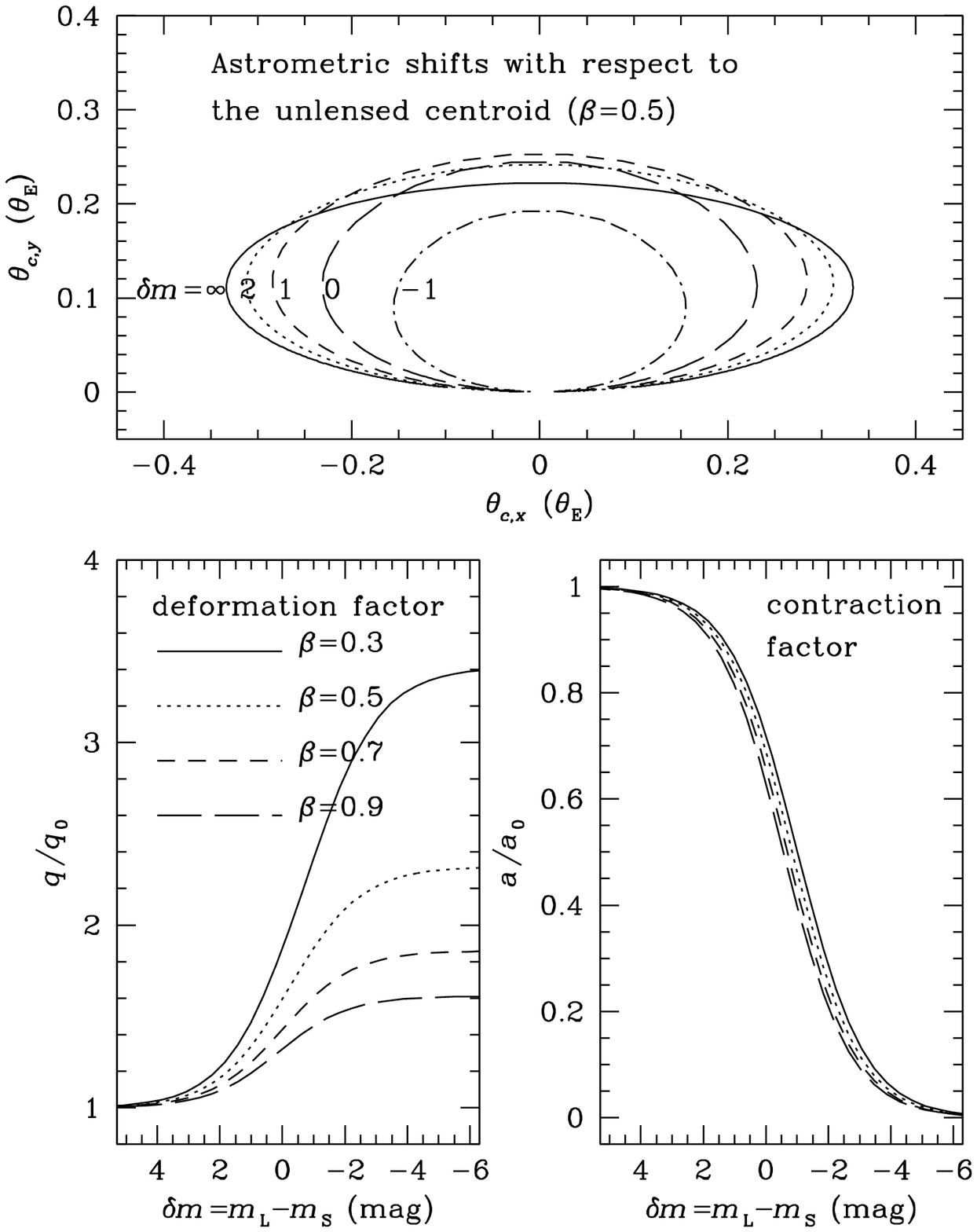}}}
\noindent
{\small {\bf Figure 2:}\
Upper panel:
The trajectories of the source image centroid displacements of events 
caused by bright lenses.  We present the trajectories for various 
values of magnitude difference between the lens and the source star, 
$\delta m$, for an example event with $\beta=0.5$.  
Lower panel: 
	The changes in the axis ratio ($q/q_0$: deformation factor)
	and in the semimajor axis ($a/a_0$: contraction factor) of the 
	astrometric ellipse of bright lens event as a function of 
	$\delta m$.
}
\end{figure}

\end{document}